\documentclass[aps,twocolumn,floatfix,showpacs]{revtex4}
\usepackage{graphicx}
\usepackage{amsmath}
\usepackage{amssymb}
\usepackage{bm}

\begin{document}

\title{Superslow Self-Organized Motions in a Multimode Microwave \\
Phonon Laser (Phaser) under Resonant Destabilization \\ of
Stationary Acoustic Stimulated Emission}
\author{D.~N.~Makovetskii}%
\email{makov@ire.kharkov.ua} \affiliation{Usikov Institute of
Radiophysics and Electronics of National Academy of Sciences, \\
12, Ac.~Proskura street, Kharkov, 61085, Ukraine}

\date{February 6, 2004}

\begin{abstract}
Two qualitatively different kinds of resonant destabilization of
phonon stimulated emission (SE) are experimentally revealed for
periodically forced multimode ruby phaser (phonon laser) operating
at microwave acoustic frequencies $\Omega_N \approx 9$~GHz
(acoustic SE wavelength $\approx 1\mu m$). The inversion state of
${\mathrm{Cr}}^{3+}$ electronic spin-system in ruby was created by
electromagnetic pump with frequency $\Omega_P = 23$~GHz. Under deep
modulation of pump power at low frequencies ($\omega_m =
70-200$~Hz, where $\omega_m$ is the modulation frequency) a fast
deterministic chaotic alternation of the microwave phonon SE modes
with different indexes $N$ is observed. This range of SE
destabilization corresponds to the relaxational resonance that is
well known for optical class-B lasers. Outside the relaxational
resonance range, namely at ultra-low (infrasonic) frequencies
$\omega_m \approx 10$~Hz of electromagnetic pump modulation, the
other kind of resonant destabilization of stationary phonon SE is
observed for the first time. This new nonlinear resonance (we call
it $\lambda$-resonance) manifests itself as very slow and
periodically repeated self-reconfiguration of the acoustic
microwave power spectra (AMPS) of a phaser generation. The period
of such self-organized motions depends significantly on $\omega_m$
and changes by several orders of magnitude when $\omega_m$ varies
within several percent. Near the vertex of $\lambda$-resonance the
period of AMPS self-reconfigurations takes giant values of several
hours (at $T=1.8$~K). The second kind of SE resonant
destabilization is explained in terms of antiphase energy exchange
between acoustic SE modes in a modulated phaser. The role of
polarized nuclear spin-system (formed by ${\mathrm{Al}}^{27}$
nuclei of the ruby crystalline matrix
${\mathrm{Al}}_2{\mathrm{O}}_3$) in these superslow self-organized
motions is discussed.


\end{abstract}

\maketitle

\section{Introduction}

The possibility of phonon stimulated emission (\mbox{SE}) in
activated crystals was considered as early as 1960s
\cite{Townes1960,Kittel1961}. Yet, speculation about various
mechanisms of phonon \mbox{SE} persists even today (see, e.g.,
\cite{Lozovik2001}). In experiments, phonon \mbox{SE} was first
observed and studied
\cite{TuckerE1961,Shiren1965,TuckerE1969,Peterson1969,DAN1974} on
dielectric crystals doped by Fe-group paramagnetic ions. The
\mbox{SE} effect manifests itself as the quantum paramagnetic
amplification of a coherent microwave phonon flux (hypersound) when
spin levels that may take part in spin-phonon interaction are
inversely populated. This effect may be viewed as an acoustic
analogue
\cite{TuckerE1961,Shiren1965,TuckerE1969,Peterson1969,DAN1974,Golenishchev1977}
of maser amplification of electromagnetic waves \cite{Siegman1966}
(not counting a number of features of nonlinear processes in the
signal and pump channels \cite{FTT1982,AR1984,ZhTF1991,RiE2003}).

At the same time, the mechanism of quantum {\textit{generation}} of
microwave acoustic oscillations, which was discovered
experimentally in \cite{TuckerE1964,SSC1974}, has long remained
unclear. The reasons were the attempts to draw analogy (see, e.g.,
\cite{Fain1982}) between acoustic quantum generators (phasers) and
usual electromagnetic quantum generators of maser type, which does
exist between related quantum amplifiers.

In experimental studies \cite{AR1984,ZhETF1977,MSMW2001} of
microwave acoustic \mbox{SE} in
${\mathrm{Ni}}^{2+}$:${\mathrm{Al}}_2{\mathrm{O}}_3$ and
${\mathrm{Cr}}^{3+}$:${\mathrm{Al}}_2{\mathrm{O}}_3$ crystals, it
was shown that operation of microwave phaser generator is
physically much closer to optical or near-infrared (near-\mbox{IR})
lasing than to operation of microwave maser generator. In fact, the
hypersonic wavelength in a Fabry-Perot acoustic resonator
(\mbox{FPAR}) is roughly 1--3~{$\mu$}m (i.e., falls into the
near-\mbox{IR} range). The quality factor $Q_C$ of an \mbox{FPAR},
as well as the quality factor of electromagnetic cavities in many
lasers, is high: $Q_C \approx 10^5 - 10^6$
\cite{AR1984,ZhETF1977,MSMW2001} (certainly, this value is reached
at liquid helium temperatures, when the non-resonant absorption of
hypersound is low). Therefore, experimental \mbox{SE} spectra of
phasers operating in the autonomous regime \cite{AR1984,SSC1974,
ZhETF1977} sometimes are similar to those observed for class-B
multimode solid-state lasers (for which $\tau_1 \gg \tau_C \gg
\tau_2$, where $\tau_1$ and $\tau_2$ are the longitudinal and
transverse relaxation times for active centers, $\tau_C$ is the
lifetime of field excitations in the cavity).

However, phasers differ radically from lasers in regard to the
intrinsic quantum noise (spontaneous emission) intensity
$J_{\mathrm{spont}}$. Since the velocity of hypersound $v_u$ is
five orders of magnitude lower than the velocity of light, the
\mbox{SE} frequency $\Omega$ in a phaser with a hypersonic
wavelength of 1--3~{$\mu$}m lies in the range $\Omega \approx 3 -
10$~GHz \cite{AR1984,ZhETF1977,MSMW2001}, i.e., is five orders of
magnitude lower than in a laser. Accordingly, the relative
spontaneous emission intensity in a phaser is $\approx 15$ orders
of magnitude lower than in a laser (because $J_{\mathrm{spont}}$
grows as $\Omega^3$). In essence, a phaser may be considered as a
{\textit{deterministic}} dynamic system throughout the \mbox{SE}
intensity range available. This is of crucial importance for
studying motion in systems with a complex stratified phase space.
It is known that multiplicative noise (including spontaneous
emission in a nonlinear active medium) affects the behavior of
dynamic systems in a very intriguing manner \cite{Horsthemke1987},
causing coarsening of the phase space topology \cite{Pecora1991},
etc.

Earlier \cite{UFZh1998}, we experimentally revealed strong dynamic
narrowing of the \mbox{SE} spectra in a nonlinear \mbox{FPAR} with
modulated electromagnetic pump of paramagnetic centers. This
narrowing was attributed to the resonant destabilization of energy
exchange between \mbox{FPAR} hypersonic (microwave acoustic) modes
under ultra-low-frequency (infrasonic) forcing of ruby
(${\mathrm{Cr}}^{3+}$:${\mathrm{Al}}_2{\mathrm{O}}_3$) spin-phonon
system. Further investigation of this microwave deterministic
system as a part of nonautonomous phaser allowed us to reveal its
still more unexpected property: superslow large-scale laminar
self-reconfigurations of the \mbox{SE} spectra akin to autowave
motions \cite{MSMW2001,PVZhTF2001}. Below, we report these
experimental studies in details and discuss the nature of the
phenomena observed.

\section{Phaser System}

\subsection{Microwave Fabry-Perot Acoustic Resonator, Active Centers,
and Hypersonic Transducer}

Experiments were carried out by using a ruby phaser
\cite{AR1984,UFZh1998,RiE1999} with the electromagnetic pump power
$P$ periodically modulated at ultra-low (infrasonic) and low (usual
sonic) frequencies: $\omega_m/2\pi = 1$~Hz -- $3$~kHz (hereafter,
the factor $2\pi$ will be omitted). A solid-state \mbox{FPAR},
which was made of synthetic single-crystal pink ruby, had the form
of a cylinder with a diameter $d_C = 2.6$~mm and length $L_C =
17.6$~mm. The end faces of the cylinder are parallel to each other
and optically flat: they serve as acoustic mirrors for hypersonic
waves. The 3-rd order crystallographic axis $\vec{\mathcal{O}}_3$
of the ruby coincides with the geometrical axis
$\vec{\mathcal{O}}_C$ of the \mbox{FPAR}. The concentration of
${\mathrm{Cr}}^{3+}$ ions is $C_a = 1.3 \cdot 10^{19}$
~${\mathrm{cm}}^{-3}$ (i.e. $\approx 0,03$\%). All measurements
were made in the interval of temperatures $T = 1.8 - 4.2$~K.

For a hypersonic frequency near $\Omega = 9.1$~GHz and $L_C$
mentioned above, the separation between longitudinal acoustic modes
of the \mbox{FPAR} is $\Delta\Omega_N^{(0)} \equiv \Omega_N^{(0)} -
\Omega_{N - 1}^{(0)} = 310$~kHz. Here, $\Omega_N^{(0)}$ is the
frequency of an $N$-th mode of the \mbox{FPAR} in the ``cold''
regime, i.e., at $P = 0$. The frequencies of the hypersound
emitted, i.e., the frequencies of phonon \mbox{SE} modes $\Omega_N
\approx \Omega_N^{(P)}$ in the ``hot'' regime ($P > P_g$), where
$P_g$ is the pump power at which phaser generation starts, lie near
9.12~GHz according to the frequency $\Omega_S = \Omega_{32} \equiv
\hbar^{-1}\left[E_3(H) - E_2(H)\right]$ of the inverted spin
transition $E_3 \leftrightarrow E_2$ of active centers in a static
magnetic field $H \approx H_0$. Thus, the frequency $\Omega_S$
corresponds to the vertex of the acoustic paramagnetic resonance
(\mbox{APR}) line coinciding by magnetic field with two electron
spin resonance (\mbox{ESR}) lines allowed for the electromagnetic
pump (Fig.~1).

The value of $H_0$ depends on the frequency $\Omega_P$ of the pump
microwave field, which saturates the spin transitions $E_1
\leftrightarrow E_3$ and $E_2 \leftrightarrow E_4$, with $\Omega_P
= \Omega_{31} = \Omega_{42} \gg \Omega_S$, where $\Omega_{31}
\equiv \hbar^{-1}\left[E_3(H_0) - E_1(H_0)\right]$; $\Omega_{42}
\equiv \hbar^{-1}\left[E_4(H_0) - E_2(H_0)\right]$ (see. Fig.~1).
The symbols $| \psi_i \rangle$ denote wave functions that belong to
the energy levels $E_i$ of the ground spin quadruplet (spin $S =
3/2$, orbital quantum number $L = 0$) of a ${\mathrm{Cr}}^{3+}$ ion
in the crystal field of ruby. Since this field is of trigonal
symmetry, all $E_i$ and $| \psi_i \rangle$ depend only on $H \equiv
|\vec{H}|$ and azimuthal angle $\vartheta$ between $\vec{H}$ and
$\vec{\mathcal{O}}_3$ \cite{Siegman1966}.

\begin{figure}
  \centering
  \includegraphics{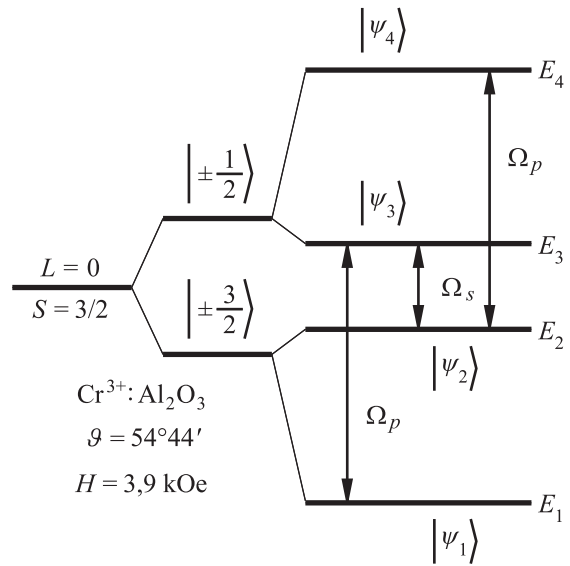}
  \caption{ \label{fg:01}%
Push-pull scheme of the energy levels of $\mathrm{Cr^{3+}}$ active
centers in pink ruby. The lowest spin quadruplet (orbital quantum
number $L=0$, spin quantum number $S=3/2$) is splitted by electric
crystal field into two doublets $\left | \pm\frac{1}{2} \right
\rangle$ and $\left |  \pm\frac{3}{2} \right \rangle$. Static
magnetic field $\vec{H}$ splits these doublets giving the symmetric
scheme of energy levels at $\vartheta = 54^{\circ}44^{\prime}$,
were $\vartheta$ is the angle between $\vec{H}$ and ruby
crystallographic axis $\vec{\mathcal{O}}_3$. Energy levels $E_i$
and wave functions $\left | \psi_i \right \rangle$ are the
eigenvalues and eigenfunctions of the ${\mathrm{Cr^{3+} :
Al_{2}O_{3}}}$ spin-hamiltonian (see e.g. \cite{Siegman1966}). At
$H \approx 3.9$~kOe the push-pull pump frequency $\Omega_P \equiv
(E_4 - E_2)/\hbar \equiv (E_3 - E_1)/\hbar \approx 23$~GHz and the
signal frequency $\Omega_S \equiv (E_3 - E_2)/\hbar \approx 9$~GHz.
}
\end{figure}

One of the \mbox{FPAR} mirrors was covered by a thin (approximately
0.5~{$\mu$}m thick) textured ${\mathrm{ZnO}}$ piezoelectric film
with a 0.1~{$\mu$}m-thick ${\mathrm{Al}}$ sublayer (both layers
were applied by evaporation in vacuum). The texture axis runs
normally to the \mbox{FPAR} mirror. The ${\mathrm{ZnO}}$ film is
the basic component of a bidirectional hypersonic transducer
designed for converting a microwave phonon field to an
electromagnetic field and vice versa. The phonon \mbox{SE} arising
in the \mbox{FPAR} excites electromagnetic oscillation in the
${\mathrm{ZnO}}$ film, and the electromagnetic signal may be
detected by standard microwave techniques. On the other hand,
exciting the ${\mathrm{ZnO}}$ film from the outside by
electromagnetic waves with a frequency $\Omega_S$, we inject
hypersonic waves with the same frequency into the \mbox{FPAR}, with
$\lambda_u^{(S)} \approx 3.3 \cdot 10^{-4}\lambda_e^{(S)}$, where
$\lambda_u^{(S)}$ is the wavelength of longitudinal hypersound in
our system, $\lambda_u^{(S)} \approx 1$~{$\mu$}m; and
$\lambda_e^{(S)} \approx 3$~cm is the wavelength of an
electromagnetic wave of the same frequency as the hypersound wave).

\subsection{Inversion States of Active Centers
and Conditions of Phaser Generation}

As it was noted earlier, inverted spin states of $E_3
\leftrightarrow E_2$ transition of ${\mathrm{Cr}}^{3+}$ active
centers are formed by the pump microwave electromagnetic field. The
frequency $\Omega_P$ of the pump source may be tuned within $22 -
25$~GHz; that is, the pump wavelength $\lambda_e^{(P)}$ belongs to
1.25~cm range of microwaves. Maximal pump power at resonant
frequency $\Omega_P \approx 23$~GHz reaches $P =
P^{({\mathrm{full}})} = 12$~mW. Through a diffraction coupler, the
pump field is excited in the cylindrical electromagnetic pump
cavity of type $H_{011}$, which has an eigenfrequency 23.0~GHz,
quality factor $Q_{CP} \approx 8 \cdot 10^3$, and geometrical
length coinciding with the length $L_C$ of the \mbox{FPAR}.

The ruby \mbox{FPAR} is placed in the pump cavity along its axis.
If $P = 0$ and the magnitude and direction of $\vec{H}$ are beyond
the \mbox{APR} range, the absorption of the hypersound injected
into the \mbox{FPAR} depends on such the two parameters:

\vspace{6pt}

($i$) the nonresonant volume attenuation $\eta_{\mathrm{vol}}$
(including losses on the lateral surfaces of the \mbox{FPAR}) and

($ii$) losses on the \mbox{FPAR} mirrors $\eta_{\mathrm{mirr}}$.

\vspace{6pt}

If $P = 0$ and the magnitude and direction of $\vec{H}$ fall into
the \mbox{APR} range (i.e., $\vec{H} \approx \vec{H_0}$), the third
absorption mechanism comes into play:

\vspace{6pt}

($iii$) the resonant paramagnetic absorption of the hypersound,
which depends considerably on the frequency of the signal injected
and on the offset of the magnetic field from the APR line vertex
\cite{Altshuler1972}.

\vspace{6pt}

 Finally, $\vec{H} \approx \vec{H_0}$ and the pump
power is applied. Then, the resonant paramagnetic absorption of the
hypersound decreases. If, as $P$ rises, one succeeds in passing
into the range where the paramagnetic absorption becomes
{\textit{negative}} (i.e., the inversion ratio $K(P, \vec{H})$
becomes {\textit{positive}}), nonparamagnetic losses of the
hypersound in the \mbox{FPAR} are compensated for partially or
completely. The complete compensation of the losses (i.e., the
onset of phaser generation) takes place first at that mode (let its
frequency be $\Omega_1$) closest to the center of the \mbox{APR}
inverted line, for which the condition
\begin{equation}
  \frac{1}{Q_{\mathrm{vol}}^{(1)}} +
  \frac{1}{Q_{\mathrm{mirr}}^{(1)}} +
  \frac{1}{Q_{\mathrm{magn}}^{(1)}} < 0,
  \label{eq:01}
\end{equation}
is met prior to other modes. In (\ref{eq:01}),
$Q_{\mathrm{vol}}^{(1)} = k_1 / \eta_{\mathrm{vol}}$,
$Q_{\mathrm{mirr}}^{(1)} = k_1 / \eta_{\mathrm{mirr}}$, $k_1 =
\Omega_1 / v_u$, and $Q_{\mathrm{magn}}^{(1)}$ is the negative (at
$K > 0$) magnetic ``quality factor'' of this mode (for which phaser
generation starts first). This magnetic ``quality factor'' is given
by
\begin{equation}
  Q_{\mathrm{magn}}^{(1)} =
  \frac{- k_1}{\alpha_1 (P, \vec{H}, \Omega_1)} \equiv
  \frac{- k_1}{K(P, \vec{H}) \sigma(\vec{H}, \Omega_1)}
  \label{eq:02}
\end{equation}
where $\alpha_1$ is the hypersound positive (at $K(P) > 0$) quantum
gain (increment of amplification) for the mode under consideration
and $\sigma$ is the hypersound paramagnetic absorption at $P=0$
(i.e at $K(0)=-1$).

The expression for $\sigma$ has the form (see, e.g., page 283 in
book \cite{Altshuler1972}):
\begin{equation}
  \sigma_{mn} =
  \frac{ 2 {\pi}^2 C_a {\nu}^2 g(\nu) |\Phi_{mn}|^2  }%
  {(2S + 1) {\rho}^{\prime} v_u^3 k_B T },
  \label{eq:03}
\end{equation}
where $\nu = \Omega / 2 \pi$, $g(\nu)$ is the form factor of the
\mbox{APR} line, ${\rho}^{\prime}$ is the crystal density, $k_B$ is
the Boltzmann constant, and the matrix element $\Phi_{mn}$
characterizes coupling of an $E_m \leftrightarrow E_n$ spin
transition with a resonant hypersonic wave of given propagation
direction and polarization.

The form factor of the \mbox{APR} line is normalized to unity,
\begin{equation}
  \int_0^{\infty} g(\nu)d{\nu} = 1,
  \label{eq:04}
\end{equation}
and the matrix element $\Phi_{mn}=\partial\langle \psi_m |
\hat{\mathcal{H}} | \psi_n \rangle / {\partial \varepsilon_{zz}}$
for the longitudinal hypersonic wave traveling along the ruby
crystallographic axis $\vec{\mathcal{O}}_3$  (the axis
$\vec{\mathcal{O}}_3$ is aligned with the $z$ coordinate axis) is
given by:
\begin{equation}
  \Phi_{mn}  = \frac{G_{33}}{2} \bigl(
  3 \langle \psi_m | \hat{S}_z^2 | \psi_n \rangle
  - S(S+1) \langle \psi_m | \psi_n \rangle
  \bigr),
  \label{eq:05}
\end{equation}
where $\varepsilon_{zz}$ is the component of the elastic strain
tensor, $\hat{\mathcal{H}}$ is the Hamiltonian of spin-phonon
interaction \cite{TuckerE1969,Golenishchev1977}, ${G_{33}}$ is the
component of the spin-phonon interaction tensor (in the Voigt
pair-index form \cite{TuckerE1969}), and $\hat{S}_z$ is the
projection of the vectorial spin operator on the $z$ axis.

To estimate $\Phi_{mn}$ we use the value $G_{33} =
5.8~{\mathrm{cm}}^{-1}$ = $1.16 \cdot 10^{-15}$~erg (found by us
experimentally by means of direct \mbox{APR} measurements at $T =
1.8 - 4.2$~K) and the wave functions $| \psi_i \rangle$ for
${\mathrm{Cr}}^{3+}$ ion in ruby (calculated in \cite{Siegman1966}
from experimental \mbox{ESR} data). With $H = 3,92$~kOe and
$\vec{H}$ directed at the angle $\vartheta =
\vartheta_{\mathrm{symm}}$ to the $z$ axis, where
$\vartheta_{\mathrm{symm}} \equiv \arccos(1/\sqrt{3}) =
54^\circ44'$, we find from (\ref{eq:05}) that $\Phi_{mn} \approx
10^{-15}~{\mathrm{erg}}$. The choice $\vartheta =
\vartheta_{\mathrm{symm}}$ refers to the so-called symmetric (or
push-pull) pump conditions \cite{Siegman1966}. Such conditions are
set up owing to the equality $E_4 - E_2 = E_3 - E_1$ (Fig.1), which
takes place at $\vartheta = \vartheta_{\mathrm{symm}}$ and provides
the most efficient inversion at the signal transition $E_3
\leftrightarrow E_2$ in the spin system. Eventually, with $\nu_S =
9.1~{\mathrm{GHz}}$, $g(\nu_S) = 10^{-8}~{\mathrm{s}}$, $C_a = 1.3
\cdot 10^{19}~{\mathrm{cm}}^{-3}$, $\rho^{\prime} =
4~{\mathrm{g/cm}}^3$, $v_u = 1.1 \cdot 10^6~{\mathrm{cm/s}}$, $T =
1.8~{\mathrm{K}}$ we find from (\ref{eq:03}) that $\sigma \approx
0.04~{\mathrm{cm}}^{-1}$.

The loaded acoustic quality factor $Q_C^{(0)}$ of the ruby
\mbox{FPAR} (with the piezoelectric film on one of the acoustic
mirrors) was measured at $T = 1.8$~K by the pulsed echo method at
frequencies $\Omega = 9.0 - 9.2$~GHz. With $\vec{H} = 0$ and $P =
0$, it was found that $Q_C^{(0)} \approx (5.2 \pm 0.4) \cdot 10^5$
for all longitudinal acoustic modes falling into this frequency
interval. Hence, $\eta \equiv \eta_{\mathrm{vol}}$ +
$\eta_{\mathrm{mirr}} = \Omega [Q_C^{(0)}v_u]^{-1} \approx
0.1~{\mathrm{cm}}^{-1}$.

The parameters $\sigma$, $\eta$, and $\alpha = \alpha_g$ (here
$\alpha = \alpha_g$ is the value of $\alpha$ at which phaser begins
generation) are obviously related as
\begin{equation}
  \alpha_g = \eta = K_g\sigma,
  \label{eq:06}
\end{equation}
where $K_g$ is the critical value of the inversion ratio $K$ for
the transition $E_3 \leftrightarrow E_2$.

Substituting $\sigma \approx 0.04~{\mathrm{cm}}^{-1}$ and $\eta
\approx 0.1~{\mathrm{cm}}^{-1}$ into (\ref{eq:06}) yields $K_g
\approx 2.5$. This value is readily attained in the case of
push-pull pump, which provides $K_{\mathrm{max}} \approx 3.3$ under
the conditions of our experiments.

\section{Experimental Results}
\label{sec:03}

\subsection{Regime of Free Generation in Ruby Phaser}

Since the frequency width $\Gamma_{32}$ of the \mbox{APR} line at
the spin transition $E_3 \leftrightarrow E_2$ is
$\approx$~$100$~MHz and the mode separation is as small as $\approx
300$~kHz, single-mode \mbox{SE} changes to multimode one even if
the pump threshold is exceeded slightly. For $\Omega_P =
\Omega_{CP}^{(0)} = 23.0$~GHz and $H = H_0 = 3.92$~kOe, free
multimode phaser generation is observed even at $P \ge
50$~{$\mu$}W. If $\Delta_H \equiv H - H_0 \neq 0$, the pump power
must be much higher (by one to two orders of magnitude) for the
condition $K > K_g$ to be fulfilled.

With pump switched on stepwise, the free phaser generation begins
in a damped oscillatory regime. For our system, the frequency
$\omega_R$ of these damped oscillations (the so-called relaxation
frequency \cite{Tang1963}) lies in the low-frequency range,
$\omega_R \approx 130$~Hz at $H = H_0$ \cite{ZhTF1989,RiE2001}.

In a free-running multimode phaser, the number of acoustic modes in
acoustic microwave power spectra (\mbox{AMPS}) does not exceed
about thirty even if $P = P^{({\mathrm{full}})} \gg P_g$. That is,
the maximal width of \mbox{AMPS} for autonomous phaser generator
($30 \times 310$~kHz $\approx 10$~MHz) is one order of magnitude
smaller than $\Gamma_{32}$, which is explained by the well-known
Tang-Statz-deMars mechanism (exhaustion of power supplies for
spatially-overlapping competing modes of \mbox{SE})
\cite{Tang1963}. If magnetic field offset is absent, free phaser
generation proceeds under near-steady-state conditions (the
integral intensity $J_\Sigma$ of multimode \mbox{SE}, which is
registered by the hypersonic transducer on one of the \mbox{FPAR}
mirrors, is virtually time independent).

If the offset is small, $|\Delta_H| \leq 3$~Oe, the value of
$J_\Sigma$ also remains practically time independent. Only with
$|\Delta_H|$ increasing up to $\approx 30$~Oe, the integral
intensity $J_\Sigma$ of phonon \mbox{SE} oscillates weakly because
few of the individual \mbox{SE} modes demonstrate small-scale
moving along the frequency axis (with correspondent changing of
their \mbox{SE} intensities) or decay at all \cite{MSMW2001}. At
$|\Delta_H| \geq 30$~Oe, some of the free generation modes in
\mbox{AMPS} becomes splitted.

These splittings are usually of order of several kilohertz (more
rarely several tens of kilohertz), which is much less than
$\Delta\Omega_N \approx 300$~kHz; the number of splitted modes is
one or two (three at most) even if $|\Delta_H| \approx 100$~Oe. The
peak intensities of splitted components are much lower than that of
unsplitted \mbox{SE} modes. This fine structure of \mbox{SE} modes
under free generation regime is of oscillating character, with not
only the spectral component amplitudes but also their frequency
positions varying smoothly (the latter within 10 kHz). The only
exceptions are narrow windows in sets of control parameters $\{P,
\Delta_H, \ldots\}$, where the spontaneous breaking of \mbox{SE}
coherence takes place for one or two modes, causing small-scale
phonon turbulence \cite{MSMW2001}. In most cases, however, free
microwave acoustic generation in a ruby phaser proceeds under
near-steady-state regime even if the offset by static magnetic
field is as large as $|\Delta_H| \approx 100$~Oe.

\subsection{Resonant Destabilization of Phonon Stimulated Emission
under Relaxational (Low-Frequency) Resonance}

The situation changes when the pump modulation frequency lies near
$\omega_m \approx \omega_R \approx 10^2$~Hz, where pronounced
nonlinear low-frequency resonance is observed
\cite{ZhTF1989,RiE2001}. If the depth of periodic modulation is
small, the phonon \mbox{SE} integral intensity $J_{\Sigma}(t)$
oscillates synchronously with the external force period:
$J_{\Sigma}(t) = J_{\Sigma}(t+\tau_m)$, where $\tau_m = 2\pi /
\omega_m$. As the pump modulation depth $k_m$ increases, the period
of $J_{\Sigma}(t)$ doubles according to the Feigenbaum scenario
$J_{\Sigma}(t) = J_{\Sigma}(t+2^f\tau_m)$, where $f = f(k_m)$
successively takes values $f = 1,2,3,...$, which condense ($f
\rightarrow \infty$) in the vicinity of a critical point $k_m =
k_m^{\mathrm{(cr)}}$.

A further increase in the depth of modulation ($k_m >
k_m^{\mathrm{(cr)}}$) switches the phaser into the state of
deterministic chaos \cite{ZhTF1989,RiE2001}. In the case of hard
excitation (for example, by a pulse of hypersound injected into the
\mbox{FPAR} from the outside), a phaser with periodically modulated
pump exhibits multistability of microwave acoustic \mbox{SE}
(branching of periodic and/or deterministic chaotic regimes of the
phaser generation, which may accompanied by hysteresis)
\cite{ZhTF1989,RiE2001}. Finally, the collision of a strange
attractor with an unstable manifold that separates the upper
periodic \mbox{SE} branch leads to so-called crises (sharp changes
of the basin of a strange attractor, which are accompanied by
attractor reconfiguration) \cite{ZhTF1989,RiE2001}.

However, all the above phenomena were detected in
\cite{ZhTF1989,RiE2001} by measuring $J_{\Sigma}(t)$. More detailed
information about phaser destabilization by a periodic force can be
extracted from the microwave spectral characteristics of phonon
\mbox{SE}. It has been found (Fig.2) that, when the depth of
modulation increases, modes of \mbox{AMPS} alternate in a stepwise
manner and cover the entire spectrum of phaser generation (for the
Fig.2 in the journal variants of this paper see \cite{ZhTF2004}).

A panoramic view of the \mbox{AMPS} shows that the position of each
individual \mbox{SE} mode is apparently the same at all the stages
of the \mbox{AMPS} alternations. In  other words, an individual
\mbox{AMPS} mode may generate or not (at this or that stage of
\mbox{SE} evolution), but the position of every generating phonon
mode is practically same, as the position of corresponding ``cold''
(unpumped) \mbox{FPAR} absorptive mode. Strictly speaking, one can
see small amplitude and frequency motions of \mbox{SE} modes if the
sweep range of the spectrum analyzer is decreased by two or three
orders of magnitude, but these small-scale motions are insufficient
at the background of global chaotic reconfigurations of the
\mbox{AMPS} under relaxational resonance.

The described picture of resonant destabilization of the microwave
phonon \mbox{SE} is typical for the whole range of the broad-band
relaxational resonance at $\omega_m = 70-200$~Hz. In a number of
cases, the intensity of the most powerful components of
nonautonomous \mbox{SE} exceeds the intensity of the central
\mbox{SE} component of an autonomous phaser by two orders of
magnitude. The evolution of the \mbox{SE} spectrum shown in Fig.2
(in discrete time) illuminates the actual complicated motions in
the spin-phonon system of an acoustic quantum generator, which is
shaded by the integral value $J_\Sigma(t)$. A similar evolution of
the \mbox{AMPS} was also observed in experiments, where pump was
stationary, but magnetic field $H$ was modulated at relaxational
resonance frequencies $\omega_m = 70-200$~Hz.

\begin{figure}
  \centering
  \includegraphics{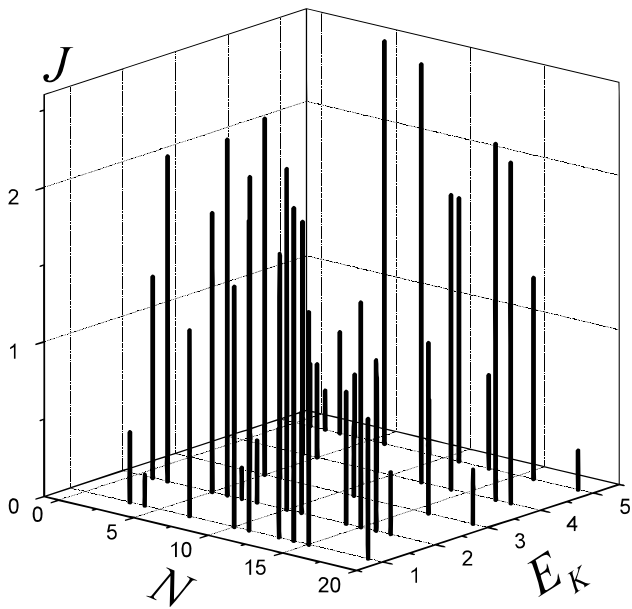}
  \caption{ \label{fg:02}
Evolution of the phonon \mbox{SE} spectra near the low-frequency
relaxational resonance at $\omega_m = 137$~Hz. The step between
sequential registered stages of \mbox{AMPS} evolution $E_K$ is
roughly equal to 1~s. }
\end{figure}

The mean lifetime of microwave phonon modes under the conditions
shown in Fig. 2 is several tenths of a second. The modes pattern
changes in an irregular manner, the mode distribution in
\mbox{AMPS} is not repeated, etc. Such a chaotic evolution of the
phonon \mbox{SE} spectra took place over the entire frequency range
$\omega_m = 70 - 200$~Hz, where resonant low-frequency
destabilization of phaser generation and, accordingly, chaotic
oscillations of $J_\Sigma(t)$ were previously observed
\cite{ZhTF1989,RiE2001}.

\subsection{Superslow Self-Organized Motions in Phaser
under Fundamental $\lambda$-Resonance}

Under the infrasonic-frequency modulation ($\omega_m \approx
10$~Hz) of pump or magnetic field, the destabilization of microwave
phonon \mbox{SE} assumes another, laminar, character. Unlike the
case of relaxational resonance ($\omega_m = 70-200$~Hz), the
resonance at infrasonic frequencies $\omega_m \approx 10$~Hz is
characterized by an extremely high correlation of spectral motions.
If the modulation frequency $\omega_m$ is precisely tuned to the
vertex $\omega_{\lambda}$ of $\lambda$-resonance and the depth of
modulation is high (close to 100\%), the \mbox{AMPS} narrows
roughly fourfold and has no greater than six or seven modes of
microwave phonon \mbox{SE}.

With a small detuning $\Delta_\lambda \equiv \omega_m -
\omega_\lambda$ of pump modulation frequency $\omega_m$ with
respect to the vertex $\omega_\lambda$ of the $\lambda$-resonance,
these narrow \mbox{SE} spectra demonstrate regular
self-reconfigurations with intriguing features. It was found
experimentally that the period of the self-reconfigurations
$\tau^{(\lambda)}_d$ changes by several orders of magnitude when
$\Delta_\lambda$ varies by no more than 1~Hz. In addition, the
period $\tau^{(\lambda)}_d$ turned out to be incommensurate with
the period $\tau_m \equiv 2\pi / \omega_m$ of external force (that
is, the frequency $\omega^{(\lambda)}_d \equiv 2\pi /
\tau^{(\lambda)}_d$ generally is not a harmonic or subharmonic of
the driving force frequency $\omega_m$ ). In experiments, this
shows up as the instability of states that have rational values
$\tau^{(\lambda)}_d / \tau_m$.

The essence of the revealed self-reconfigurations of the phonon
\mbox{SE} spectra is the $\tau^{(\lambda)}_d$-periodic
unidirectional moving of the range of active (i.e. generating)
modes along the \mbox{SE}-frequency axis $\Omega$ if $\omega_m
\approx \omega_\lambda$. This range typically comprises from three
to seven microwave phonon \mbox{SE} modes.

It is noteworthy that the frequency position of each of the
\mbox{SE} modes remains nearly unchanged (if higher order dynamic
effects due to the nonstationary fine structure of the \mbox{SE}
spectra \cite{RiE1999} are disregarded). Only the position of the
spectral part with active modes changes. Thus, the ignition of new
\mbox{FPAR} modes at one side of the \mbox{AMPS} (beginning from
some discrete starting position $\Omega_N^{({\mathrm{start}})}$) is
accompanied by the extinguishing of the same number of modes at the
opposite side of the \mbox{AMPS}. Such a motion lasts until the
phaser generation ceases completely (at some discrete finishing
position $\Omega_N^{({\mathrm{finish}})}$ of the frequency axis
$\Omega$). After a relatively short period of complete absence of
phaser generation (time of refractority), the process of
\mbox{AMPS} global self-reconfiguration is repeated, starting from
the same position $\Omega_N^{({\mathrm{start}})}$ on the frequency
axis $\Omega$.

The period $\tau^{(\lambda)}_d$ of these unidirectional spectral
motions remains the same if a set of control parameters does not
change. On the screen of a spectrum analyzer, this evolution of the
microwave phonon \mbox{SE} spectrum appears as the periodic motion
of some mode cluster (a set of simultaneously generating modes,
having fixed own frequencies). Typical sequences of microwave
phonon \mbox{SE} spectra at the $\lambda$-resonance conditions
($\omega_\lambda = 9.79$~Hz, $\Delta_\lambda = - 0.23$~Hz) in the
absence of static magnetic field offset ($\Delta_H = 0$) are shown
in Fig.3.

\begin{figure}
  \centering
  \includegraphics{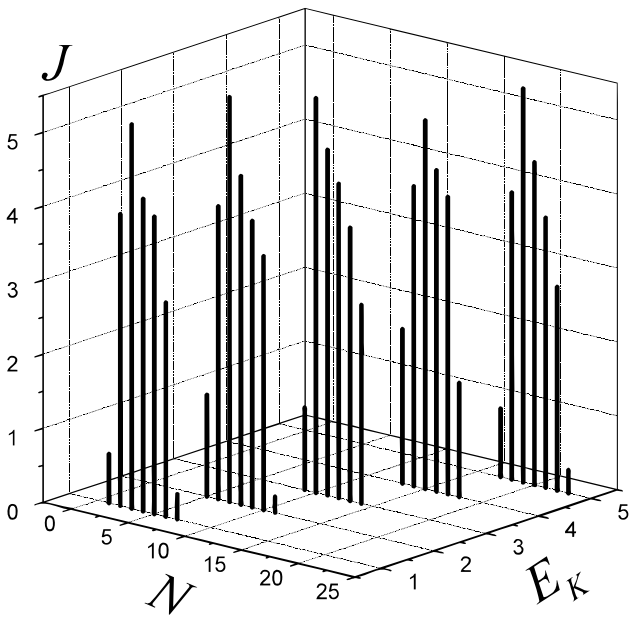}
  \caption{ \label{fg:03}%
Evolution of the phonon \mbox{SE} spectra near the ultra-low
frequency $\lambda$-resonance at $\omega_m = 9.56$~Hz. The step
between sequential registered stages of \mbox{AMPS} evolution $E_K$
is about 2.5~s. }
\end{figure}

An instantaneous set of microwave phonon \mbox{SE} modes forms a
cluster of certain width, and this width varies insignificantly
during the spectral motion (Fig.3). At the same time, a set of
{\textit{individual}} \mbox{SE} modes (that form active cluster)
permanently varies (see disrete positions $N$ of generating modes
$\Omega_N$ for successive \mbox{AMPS} snapshots $E_K$ at Fig.3). As
follows from Fig.3, such the self-reconfigurations of the
\mbox{AMPS} are imposed on irregular oscillations of the SE mode
intensity.

When the sweep range of the spectrum analyzer is decreased by two
or three orders of magnitude, very weak irregular motions of
\mbox{SE} modes along the frequency axis $\Omega $ and sometimes
splittings of these modes are observed (Fig.4). The modes split
when their intensity are very low, e.g. before extinguishing
(because of this, the instrument noise is noticeable in Fig.4).
Obviously, such fine effects cannot be seen on panoramic spectra in
Fig.3, where the positions of $\Omega_N$ are apparently fixed (the
fine structure of the \mbox{AMPS} in an autonomous ruby phaser was
studied by us in \cite{MSMW2001,RiE1999}). In general, it can be
said that large-scale ordered (laminar) motions of \mbox{SE}
spectra in a phaser with infrasonic-frequency pump modulation are
imposed on small-scale irregular motions.

Similar large-scale laminar self-reconfigurations of phonon
microwave spectra in a nonautonomous ruby phaser were observed in
experiments at $\Delta_\lambda = + 0.23$~Hz; however, the cluster
moved in the opposite direction. Further investigations showed that
the sign of the derivative $d\Omega_V/dt$ (here, $\Omega_V$ is the
center frequency of a mode cluster) strictly correlates with the
sign of frequency detuning of the external force from resonance:
${\mathrm{sgn}}[d\Omega_V/dt] = -{\mathrm{sgn}} \Delta_\lambda$.
Going toward the exact $\lambda$-resonance discovered
($|\Delta_{\lambda}| \rightarrow 0$), the self-reconfigurations
period $\tau^{(\lambda)}_d$ takes giant values. Direct measurements
of $\omega^{(\lambda)}_d \equiv 2\pi/\tau^{(\lambda)}_d$ gave
${\mathrm{inf}}(\omega^{(\lambda)}_d) < 10^{-4}$~Hz (one
self-reconfiguration period is $\approx 3$ hours!). All the
experiments on measuring of $\tau^{(\lambda)}_d$ were performed
with superfluid helium ($T = 1.8$~K) in order to avoid problems
associated with cryogenic liquid boiling.

Such a character of \mbox{AMPS} self-reconfigurations was observed
not only for zero-offset of static magnetic field ($\Delta_H = 0$);
it persists over a wide range of $\Delta_H$. Moreover, at $ |
\Delta_H | < 10$~Oe, the value of $\omega_\lambda$ even remains
almost constant (close to 9.8~Hz). Only when the magnetic field
offset modulo $ | \Delta_H | $ increases further, does the
resonance frequency $\omega_\lambda$ decrease tangibly (about
twofold for $ | \Delta_H | \approx 60$~Oe). It is essential that
the above dependence of the motion direction of an \mbox{SE} mode
cluster on the modulation frequency detuning,
${\mathrm{sgn}}[d\Omega_V/dt] = -{\mathrm{sgn}} \Delta_\lambda$,
remains valid.

\subsection{Even and Odd Harmonics of $\lambda$-Resonance}

Along with the discontinuous regimes of phaser generation (which
include the time of refractority at each period of the \mbox{AMPS}
self-reconfigurations), we also observed regimes without
refractorities. For such the regimes at least one \mbox{SE} mode
appears in the \mbox{AMPS} starting range before the last \mbox{SE}
mode in the finishing range disappears. In other words, there are
two narrow \mbox{SE} mode clusters (at the $\Omega$ axis) during
some part $\delta\tau^{(\lambda)}_d$ of the \mbox{AMPS}
self-reconfiguration period. The virtual vertici of these clusters
$V_1$ and $V_2$ are moving simultaneously along the $\Omega$ axis
in the same direction and with the same velocity: $d{\Omega_V}_1/dt
= d{\Omega_V}_2/dt$ during $\delta\tau^{(\lambda)}_d$.

The same effect was observed at the first three {\textit{even}}
harmonics of $\lambda$-resonance, i.e. at $\omega_m \approx
\omega_{2s\lambda} \equiv 2s\omega_\lambda$, where $s \in
\{1,2,3\}$. As for the fundamental $\lambda$-resonance (for which
$\omega_m = \omega_{\lambda}$), in the case of poined lowest
$2s\lambda$-resonances correspondent periods of the \mbox{AMPS}
self-reconfigurations $\tau^{(2s\lambda)}_d$ are incommensurate
with the external force period $\tau_m \equiv 2\pi / \omega_m$.
Experiments at $T = 1.8$~K show, that these periods
$\tau^{(2s\lambda)}_d$ are increasing up to 100~s and more if the
absolute values of detunings $| \Delta_L^{(2s\lambda)} | \equiv |
\omega_m - \omega_{2s\lambda}|$ are small (less than 0,05-0,1~Hz).
The sign of $d\Omega_V/dt$ (or ${\mathrm{sgn}}[d{\Omega_V}_1/dt]$,
${\mathrm{sgn}}[d{\Omega_V}_2/dt]$ for off-refractority regimes)
was opposite to the sign of $\Delta_L^{(2s\lambda)}$ (as in the
case of the fundamental $\lambda$-resonance).

With $4 \leq s \leq 11$, the driving force frequency $\omega_m$
falls into the range of broad-band relaxation resonance ($\omega_m
= 70 - 200$~Hz) described above. However, for $s > 11$ (when the
driving force leaves this range of resonant destabilization)
experiments again distinctly revealed narrow-band resonant
responses of the generating phaser system to the pump modulation.
These narrow-band resonances were observed not only at even
harmonics $\omega_{2s\lambda}$ of the fundamental
$\lambda$-resonance but also at odd harmonics
$\omega_{(2s+1)\lambda} \equiv (2s+1)\omega_{\lambda}$.

\begin{figure}
  \centering
  \includegraphics{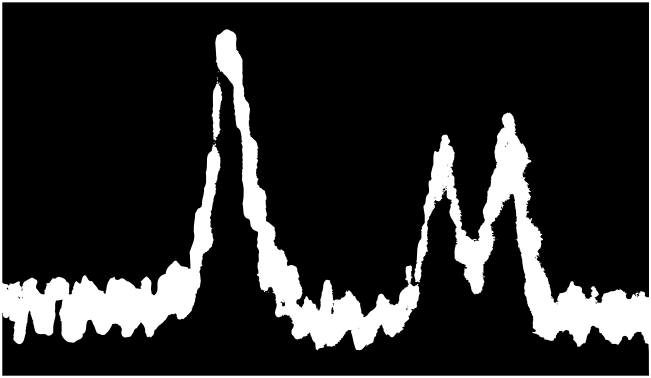}
  \caption{ \label{fg:04}
Fine structure of an individual microwave phonon \mbox{SE} mode in
the \mbox{AMPS}. The sweep along the abscissa axis is 10~kHz. }
\end{figure}

Responses of the ruby phaser at the pointed high harmonics (both
even and odd) somewhat differ from those in the case of fundamental
resonance $\lambda$-resonance and its first even harmonics. For
example, at $s > 11$, the deviation of $\Omega_V$ is, as a rule, no
greater than one or two \mbox{SE} mode separations (i.e., no
greater than 0,3--0,6~MHz). Besides of this, at $s > 11$, \mbox{SE}
modes experience deep periodic automodulation (of depth 50\% or
more).

At $s > 11$ ($\omega_m > 215$~Hz) and with detunings $\approx
1$~Hz, this automodulation is fast (its period is about a second).
As $\omega_m$ approaches the top of each of the $2s\lambda$- or
$(2s+1)\lambda$- resonances (at the same $s > 11$), the
automodulation period $\tau_{\mathrm{am}}$, as well as the period
$\tau^{(\lambda)}_d$ of the \mbox{AMPS} self-reconfigurations at
$\omega_m \approx \omega_{\lambda}$ and lower even harmonics ($s <
4$), increases monotonically. The highest values of
$\tau_{\mathrm{am}}$, which were stably observed at $11 < s < 20$,
reached several minutes for $T = 1.8$~K. Note that the same
``blinking'' regimes of phonon \mbox{SE} (but with smaller
automodulation depths) were also found for the first two
{\textit{odd}} harmonics ($\omega_m \approx 3\omega_\lambda$ and
$\omega_m \approx 5\omega_\lambda$), where ``great''
self-reconfigurations of the \mbox{AMPS} are absent.

\section{Discussion}

Our experimental data for infrasonic-frequency-modulated phaser
suggest spin-phonon self-organization caused by intermode energy
exchange at hypersonic frequencies. It should be emphasized that
under infrasonic $\lambda$-resonance highly organized collective
motions in the spin-phonon system are observed not only for each of
individual microwave acoustic modes (as in the case of autonomous
multimode phaser generation) but also at the global level, where
all SE modes obey the same rhythm (the frequency of which is not a
harmonic/subharmonic of an external perturbation). In other words,
if upon autonomous multimode phaser generation there exist $N$
virtually independent microwave oscillators (each corresponding to
a specific SE mode), at a resonant infrasonic perturbation of the
pump or magnetic field, these oscillators behave consistently
(cooperatively). The same phenomenon of global-level
self-organization in ruby phaser was observed by us at some higher
harmonics of the fundamental $\lambda$-resonance (outside of the
broad-band relaxational resonance, having low, --- but not
infrasonic, --- fundamental frequency).

The features of collective motions in a phaser suggest that this
effect is similar to antiphase dynamics processes
\cite{Georgiou1994,Nguyen1997,Vladimirov1999}, which were
discovered previously in multimode lasers. In the simple case of
two-mode lasing \cite{Georgiou1994}, antiphase dynamics appears as
consistent oscillations of modes strictly in antiphase. In $N$-mode
systems, antiphase motions may be much more complicated (see, e.g.,
\cite{Nguyen1997,Vladimirov1999}); however, the general nontrivial
tendency, namely, coherent unidirectional SE mode oscillations with
a time delay $\tau_d^{(\lambda)}/N$ between nearest neighbors,
still persists.

As was found in this work, energy exchange between microwave phonon
\mbox{SE} modes in a phaser results in an additional characteristic
frequency $\omega_\lambda$, which is much lower than the relaxation
frequency $\omega_R$. Collective motions are excited when a control
parameter (pump or magnetic field) of the active system is
modulated at frequencies close to $\omega_\lambda$. The same is
true for lasers exhibiting antiphase dynamics
\cite{Georgiou1994,Nguyen1997,Vladimirov1999}. Accordingly, phonon
spectral self-reconfigurations may be treated as the occurrence of
antiphase states in phaser system when the spatial distribution of
stationary hypersonic \mbox{SE} modes is destabilized by an
external force at infrasonic frequencies $\omega_m \approx
\omega_\lambda$ (or its harmonics). Moreover, the value of
$\omega_\lambda$ estimated by formulas given in \cite{Georgiou1994}
(see Appendix 1) is one order of magnitude lower than $\omega_R$,
which is also in agreement with our experimental data for
nonautonomous phaser generation.

It should be noted, however, that the laser model of nonlinear
dynamics of microwave phonon \mbox{SE} cannot describe adequately
all features of self-organization in a phaser near
$\lambda$-resonance and its harmonics, although it gives a
satisfactory estimate of $\omega_\lambda$ and predicts more or less
accurately the character of mode motions.

For a better understanding of self-organization in a ruby phaser,
one should consider the unusual hierarchy of spin reservoirs
\cite{Atsarkin1978}, which are responsible for many features of the
quantum transitions saturation (both electromagnetic and acoustic)
in the microwave range. Essentially, all nonlinearities showing up
in microwave resonant interactions of the signal acoustic field and
the electromagnetic pump field with the electron Zeeman reservoir
$\widetilde{Z}_E$ \cite{Atsarkin1978} of ${\mathrm{Cr^{3+}}}$ ions
in ruby phaser are sensitive to the presence of slowly relaxing
${\mathrm{Al^{27}}}$ nuclei, as revealed in our early experiments
on hypersound quantum amplification
\cite{FTT1982,AR1984,ZhETF1977}.

The reason for this sensitivity is the direct thermal interaction
\cite{Atsarkin1978} between the nuclear Zeeman reservoir
$\widetilde{Z}_N$ and electronic dipole-dipole ($d-d$) reservoir
$\widetilde{D}_E$. This interaction leads to energy exchange
between electronic and nuclear spin reservoirs: $\widetilde{Z}_E
\rightleftarrows \widetilde{D}_E \rightleftarrows \widetilde{Z}_N$.
It is important that the heat capacities of the reservoirs
$\widetilde{Z}_N$ and $\widetilde{Z}_E$ are comparable to each
other, although the frequency of nuclear magnetic resonance
(\mbox{NMR}) for ${\mathrm{Al^{27}}}$ ($\approx 10$~MHz) is three
orders of magnitude lower than the \mbox{ESR} and \mbox{APR}
frequencies for ${\mathrm{Cr^{3+}}}$ ions at $H \approx 4$~kOe.
This is because the concentration of impurity ${\mathrm{Cr^{3+}}}$
paramagnetic ions ${\mathrm{Cr^{3+}}}$ in pink ruby is as low as
several hundredths of a percent; that is, for one electron spin,
there are several thousands of nuclear spins. As a result, the
inertial nuclear system, while unseen in direct \mbox{ESR} and
\mbox{APR} measurements, participates in all population
redistributions over electron spin levels (in more exact terms,
over a quasicontinuous set of sublevels due to dipole-dipole
interactions \cite{Atsarkin1978}).

Returning back to the $\lambda$-resonance, we may assume that the
observed decrease in its frequency at static magnetic field
mismatches is caused by ${\mathrm{Al^{27}}}$ nuclear spin-system,
which is thermally connected to ${\mathrm{Cr^{3+}}}$ electronic
spin-system. More precisely, $\widetilde{Z}_N$ is involved (by the
direct thermal contact to $\widetilde{D}_E$ and by the indirect one
to $\widetilde{Z}_E$) in energy exchange between microwave acoustic
\mbox{SE} modes. In fact, the most important feature of interaction
between the saturated electron subsystem and polarized nuclear
subsystem is the strong dependence of the spin temperatures (in all
the three reservoirs $\widetilde{Z}_E$, $\widetilde{D}_E$,
$\widetilde{Z}_N$ ) on static magnetic field mismatches and/or
saturating microwave field detunings in ruby phaser (see our early
experimental works \cite{FTT1982,AR1984,ZhETF1977}).

It is also noteworthy that the effect of the low-energy (combined)
reservoir $\widetilde{Z}_N + \widetilde{D}_E$ on the high-energy
reservoir $\widetilde{Z}_E$ in a phaser differs radically from the
effect of similar low-energy reservoir in optical lasers (see,
e.g., \cite{Arecchi-1989}). The matter is that, in our system, the
relative heat capacity and inertia of combined low-energy reservoir
are much greater than in ${\mathrm{CO_2}}$ lasers
\cite{Arecchi-1989}. First, in our system, as was noted above, for
one active center ${\mathrm{Cr^{3+}}}$ there are several thousands
of magnetic nuclei ${\mathrm{Al^{27}}}$, which make a low-energy
reservoir ``heavier'', while in ${\mathrm{CO_2}}$ lasers, the low-
and high-energy reservoirs are formed by the same ${\mathrm{CO_2}}$
molecules (the low-energy reservoir is formed here by
rotational-vibrational degrees of freedom of ${\mathrm{CO_2}}$
molecules). So, in a ${\mathrm{CO_2}}$ laser, there are analogues
to the reservoirs $\widetilde{Z}_E$ and $\widetilde{D}_E$, but that
laser does not have an analogue to the reservoir $\widetilde{Z}_N$.
Second (and most important), the relaxation time of the low-energy
reservoir in ruby phaser is much longer than the relaxation time of
the high-energy reservoir \cite{FTT1982,AR1984,ZhETF1977}, while
{\textit{the reverse}} is true for ${\mathrm{CO_2}}$ laser
\cite{Arecchi-1989}. Therefore, in a ruby phaser, the combined
inertial reservoir $\widetilde{Z}_N + \widetilde{D}_E$ is involved
in the leading self-organization processes which always the slowest
processes are.

\section{Conclusions}

We experimentally studied the influence of an external periodic
force on the dynamics of microwave phonon stimulated emission in a
ruby phaser at liquid helium temperatures. It is shown that the
periodic modulation of pump (or static magnetic field) at low
frequencies $\omega_m = 70-200$~Hz chaotizes energy exchange in the
spin-phonon system because of resonant destabilization of a phaser
near its relaxation resonance frequency $\omega_R$. In this range
of destabilization, the width of the phonon microwave power
spectrum does not noticeably change. For ultra-low (infrasonic)
frequencies of pump modulation ($\omega_m \approx \omega_\lambda
\approx 10$~Hz), a qualitatively new kind of phonon \mbox{SE}
destabilization is discovered. First, the \mbox{AMPS} narrows
considerably (almost four times). Instead of fast chaotic \mbox{SE}
modes alternations, the self-organized periodic slow motions of
\mbox{SE} modes near the vertex of the $\lambda$-resonance are
observed. Period of these motions may exceed the period of the
driving force by several orders of magnitude. The phonon \mbox{SE}
self-organization is of a global type: cooperative behaviour in the
infrasonically-modulated phaser includes the whole spin-phonon
system generating microwave phonons (unlike conventional
``inividualistic'' intramode self-organization, which is typical of
multimode lasers).

Self-organization shows up as slow consistent regular pulsations of
each of microwave phonon \mbox{SE} modes with a time delay
$\tau^{(\lambda)}_d / N$. This appears as autowave motion of a mode
cluster in the spectral space. The total self-reconfiguration cycle
of \mbox{AMPS} depends considerably on $\omega_m - \omega_\lambda$
and changes by several orders of magnitude when $| \omega_m -
\omega_\lambda |$ changes by several percent. The same processes
were observed for the first three even harmonics of the fundamental
$\lambda$-resonance. For higher even harmonics, as well as for all
odd harmonics, ``blinking'' periodic regimes are discovered
(outside of wide-band low-frequency relaxational resonance). The
results obtained are treated in terms of the antiphase dynamics of
phonon \mbox{SE}. The role of the ${\mathrm{Al^{27}}}$ magnetic
nuclear subsystem (witn MHz-range \mbox{NMR}-frequencies) in
self-organization of GHz-range phonon \mbox{SE} is discussed.

\begin{acknowledgments}

The author is grateful to E.~D.~Makovetsky (Karazin National
University, Kharkov) and S.~D.~Makovetskiy (Kharkov National
University of Radioelectronics) for their valuable help with
computer processing of the experimental data, to A.~P.~Korolyuk
(Institute of Radiophysics and Electronics of \mbox{NASU}) for
interest in the study of phaser dynamics, and to P.~Mandel
(Universit\'e Libre de Bruxelles) for the kindly submitted
publications on antiphase dynamics. Finally, the author is indebted
to Academician V.~M.~Yakovenko and all participants of his seminar
for discussion and valuable comments on nonlinear phenomena in
phaser system.

This work was partially supported by the Scientific and Technology
Center of Ukraine (STCU).
\end{acknowledgments}

\appendix
\section{Nonlinear Resonances in a Two-Mode Class-B Quantum Generator}

In this Appendix, we reproduce some results of the work
\cite{Georgiou1994} on antiphase dynamics, where the nature of
additional nonlinear resonance in a simple two-mode system was
considered basing on the Tang-Statz-DeMars approximation
\cite{Tang1963}.

Equations of motion for a two-mode class-B quantum generator (with
$\tau_1 \gg \tau_C \gg \tau_2$) may be formulated using 5D
vectorial order parameter $\vec{B}$:
\begin{equation}
  \vec{B}(t) = (J_1, J_2, D_0, D_1, D_2),
  \label{eq:07}
\end{equation}
where $J_n(t)$ is the normalized dimensionless intensity of
\mbox{SE} for $n$-th mode ($n = 1; 2$); $D_j(t)$ is the spatial
Fourier components for inversion in an active medium ($j = 0; 1;
2$):
\begin{equation}
  D_j(t) = \frac{1}{L_C}\int_0^{L_C}D(z,t)\cos(2k_jz)dz.
  \label{eq:08}
\end{equation}
Here $D(z,t)$ is normalized spatio-temporal distribution of
inversion along axis ${\mathcal{O}_z}$ of the Fabry-Perot
resonator; $k_0 = 0$; $k_{1,2}$ are the wave numbers for the
\mbox{SE} modes ($k_{1,2}\gg L_C^{-1}$); $L_C$ is the length of the
Fabry-Perot resonator. The gain for the 2-nd \mbox{SE} mode
$\alpha_2$ is assumed to be less than the gain of the 1-st
\mbox{SE} mode $\alpha_1$.

Nonlinear dynamical model for a class-B quantum generator has such
the form:
\begin{equation}
  \tau_1\frac{{\mathrm{d}}\vec{B}}{{\mathrm{d}}t} =
  \vec{\Psi}^{(L)}(\vec{B}) + \vec{\Psi}^{(NL)}(\vec{B})
  \label{eq:09}
\end{equation}
where $\vec{\Psi}^{(L)}$ are $\vec{\Psi}^{(NL)}$ linear and
nonlinear parts of the vector field with the following components:

\vspace{6pt}

\begin{equation}
\left.
\begin{array}{l}
  \Psi_{1,2}^{(L)} = - 2J_{1,2}/q_1;
  \hspace{3mm} \Psi_3^{(L)} = A - D_0;
  \vspace{3mm} \\
  \Psi_{4,5}^{(L)} = - D_{1,2}; \vspace{3mm} \\
  \Psi_{1,2}^{(NL)} = (2D_0 - D_{1,2})
  \mu_{1,2}J_{1,2}/q_1; \vspace{3mm} \\
  \Psi_3^{(NL)} = [(D_1 - 2D_0)\mu_1J_1 +
  (D_2 - 2D_0)\mu_2J_2]/2; \vspace{3mm} \\
  \Psi_{4,5}^{(NL)} = \mu_{1,2}D_0J_{1,2} -
  (\mu_1J_1 + \mu_2J_2)D_{1,2}
\end{array}
\right\}
  \label{eq:10}
\end{equation}

\vspace{6pt}

\noindent where $q_1 \equiv 2\tau_C/\tau_1$, $\mu_n \equiv
\alpha_n/{\mathrm{max}}(\alpha_n)$, i.e. $\mu_1 = 1$, $\mu_2 =
\alpha_2/\alpha_1 < 1$.

Because of $q_1 \ll 1$ (in our experiments with acoustic quantum
generator $q_1 \approx 10^{-4} \div 10^{-5}$, see
Section~\ref{sec:03}), one can find from equations (\ref{eq:09}),
(\ref{eq:10}) two resonances in our active system.

The first nonlinear resonance is usual relaxational resonance with
$\omega_R = |{\mathrm{Im}}(\Lambda_{1,2})| \approx [{(4 -
D_0^{(st)})} J_1^\prime]^{1/2}$, $J_{1,2}^\prime =
J_{1,2}^{(st)}/2\tau_1\tau_C$, $M_2 = [4(1 + \mu_2) -
3\mu_2D_0^{(st)}]$, where $\Lambda_{1,2}$ is the first pair of
complex conjugate Liapunov exponents; index ($st$) corresponds to
the stationary solutions of (\ref{eq:09}). Such the resonance
presents in single-mode class-B laser too.

The second nonlinear resonance is due to intermode energy exchange
(it is obviously absent in a single-mode class-B laser). The
frequency of second nonlinear resonance is \cite{Georgiou1994}:
$\omega_{\lambda} = |{\mathrm{Im}}(\Lambda_{3,4})|$, where
\begin{equation}
  |{\mathrm{Im}}(\Lambda_{3,4})|^2 \approx J_2^\prime\frac{M_2\mu_2D_0^{(st)} -
  4{(1 - \mu_2)}^2}{4 - D_0^{(st)}} \ll |{\mathrm{Im}}(\Lambda_{1,2})|^2.
  \label{eq:11}
\end{equation}
Here $A$ is the pump parameter (normalized by the threshold of the
single-mode generation), $A > A^{(2m)}$, $A^{(2m)}$ is the
threshold of the two-mode generation:
\begin{equation}
  A^{(2m)} = \frac{4\mu_2 - 2\mu_2^2 - 1}{(2\mu_2 - 1)\mu_2}.
  \label{eq:12}
\end{equation}
Thus, intermode energy exchange in class-B quantum generator leads
to appearance of additional resonant frequency $\omega_\lambda =
|{\mathrm{Im}}(\Lambda_{3,4})|$, which is much less than $\omega_R$
and may be attributed to infrasound-frequency $\lambda$-resonance
in ruby phaser (Section~\ref{sec:03}).

\section{List of Abbreviations}

\begin{tabbing}
 \hspace{5ex}\= \hspace{8ex}\= \hspace{2ex}\= \kill
   \> AMPS \> - \> Acoustic Microwave Power Spectrum\\
   \> APR  \> - \> Acoustic Paramagnetic Resonance\\
   \> ESR  \> - \> Electron Spin Resonance\\
   \> FPAR \> - \> Fabry-Perot Acoustic Resonator\\
   \> IR   \> - \> InfraRed\\
   \> NMR  \> - \> Nuclear Magnetic Resonance\\
   \> SE   \> - \> Stimulated Emission\\
  \end{tabbing}

\end{document}